\documentclass[twocolumn,showpacs,preprintnumbers,amsmath,amssymb,showkeys]{revtex4-1}

\usepackage{graphicx}
\usepackage{dcolumn}
\usepackage{bm}
\usepackage{euscript}
\usepackage{float}

\newcommand{\e}{\epsilon}

\newcommand{\s}{\sigma}
\newcommand{\fee}{\varphi}
\renewcommand{\t}{\tau}
\newcommand{\V}{{\EuScript{V}}}
\renewcommand{\d}{\textstyle}

\renewcommand{\L}{\check{\Lambda}}

\newcommand{\tr}{\,\mathrm{tr}}

\newcommand{\bB}{\mathbb{B}}

\newcommand{\G}{\, \check{G}}
\newcommand{\Gt}{\, \check{\tilde{G}}}
\newcommand{\A}{ \mathcal{A}}
\newcommand{\D}{\mathfrak{D}}

\begin{document}

\title{Quantum phase fluctuations and density of states in superconducting nanowires}
\author{Alexey Radkevich}
\affiliation{Moscow Institute of Physics and Technology, 141700 Dolgoprudny, Russia}
\author{Andrew G. Semenov}
\affiliation{I.E. Tamm Department of Theoretical Physics, P.N. Lebedev Physical Institute, 119991 Moscow, Russia}
\affiliation{National Research University Higher School of Economics, 101000 Moscow, Russia}
\author{Andrei D. Zaikin}
\affiliation{Institute of Nanotechnology, Karlsruhe Institute of Nanotechnology (KIT), 76021, Karlsruhe, Germany}
\affiliation{I.E. Tamm Department of Theoretical Physics, P.N. Lebedev Physical Institute, 119991 Moscow, Russia}

\date{\today}

\begin{abstract}
We argue that quantum fluctuations of the phase of the order parameter may strongly affect the electron density of states (DOS) in ultrathin superconducting
wires. We demonstrate that the effect of such fluctuations is equivalent to that of a quantum dissipative environment formed by sound-like plasma modes propagating along the wire. We derive a non-perturbative expression for the local electron DOS in superconducting nanowires which fully accounts for quantum phase fluctuations. At any non-zero temperature these fluctuations smear out the square-root singularity in DOS near the superconducting gap and
generate quasiparticle states at subgap energies. Furthermore, at sufficiently large values of the wire impedance this singularity is suppressed down to $T=0$ in which case DOS tends to zero at subgap energies and exhibits the power-law behavior above the gap. Our predictions can be directly tested in tunneling experiments with superconducting nanowires.

\end{abstract}
\maketitle

\section{\label{sec:level1}Introduction}

Fluctuations play an important role in a reduced dimension. Of particular interest are fluctuation effects in low dimensional superconducting structures \cite{AGZ,LV} in which case the system behavior can be markedly different from that in the bulk limit. For instance, it is well known that properties
of quasi-one-dimensional superconducting wires cannot be adequately described within the standard Bardeen-Cooper-Schriffer (BCS) mean field approach even if the temperature $T$ becomes
arbitrarily low. Perhaps one of the most striking low temperature features of ultrathin superconducting wires is the presence of nontrivial fluctuations of the order parameter -- the so-called quantum phase slips (QPS) \cite{AGZ}. Such quantum fluctuations correspond to temporal local suppression
of the superconducting order parameter accompanied by the phase slippage process which, in turn, generates voltage fluctuations in the system. As a result,
ultrathin superconducting wires acquire QPS-induced non-vanishing resistance down to lowest $T$ \cite{ZGOZ,GZQPS}. Subsequently this theoretical prediction  received its convincing experimental confirmation \cite{BT,Lau,Zgi08}. More recently it was predicted \cite{SZ16} that QPS can also cause non-equilibrium
(shot) voltage noise in superconducting nanowires.

The magnitude of quantum phase slip effects in such nanowires is controlled by the QPS amplitude $\gamma_{QPS} \sim (g_\xi\Delta/\xi)\exp (-ag_\xi)$,
where $\Delta$ is the superconducting order parameter and $a \sim 1$ is an unimportant numerical prefactor. The key parameter here is
dimensionless conductance $g_\xi =R_q/R_\xi$, where $R_q =2\pi/e^2 \simeq 25.8$ K$\Omega$ is the quantum resistance unit and $R_\xi$ is the normal state resistance of the wire segment of length equal to the superconducting coherence length $\xi$.
The same parameter $g_\xi$ (which is directly related to the so-called Ginzburg number in one dimension \cite{LV} as  $g_\xi \sim Gi_{1D}^{-3/2}$) controls the magnitude of small (Gaussian) fluctuations of the order parameter in superconducting nanowires. E.g., it is straightforward to demonstrate \cite{GZTAPS}
that in the presence of such fluctuations the mean field value of order parameter $\Delta$ acquires a negative correction $\Delta\to \Delta -\delta \Delta$
with $\delta \Delta  \sim \Delta /g_\xi $. Thus, by choosing
the dimensionless conductance $g_\xi $ sufficiently large one can essentially suppress both QPS effects and Gaussian fluctuations of the absolute value $|\Delta|$ in superconducting wires.

Is the condition $g_\xi \gg 1$ sufficient to disregard fluctuation effects in such wires? The answer to this question is clearly negative. The point here is that even at very large values of $g_\xi$ there remain non-vanishing fluctuations of the phase $\varphi$ of the order parameter. In the limit $g_\xi \gg 1$ such phase fluctuations are essentially decoupled from those of $|\Delta |$ being
controlled by the dimensionless parameter $g=R_q/Z_{\rm w}$, where $Z_{\rm w}=\sqrt{\mathcal{L}_{\rm kin}/C}$ is the wire impedance,
$\mathcal{L}_{\rm kin}=1/(\pi\sigma_N\Delta s)$ and $C$ are respectively the kinetic wire inductance (times length) and the geometric wire capacitance (per length), $\sigma_N$ is the normal state Drude conductance of the wire and $s$ is the wire cross section. The parameter $g$ is different from
(although not unrelated to) $g_\xi $ (e.g., $g \propto \sqrt{s}$ while $g_\xi \propto s$) and, hence, by properly choosing the system parameters one can select the wires where only phase fluctuations can play a significant role. Such kind of wires will be addressed below in this paper.

To be specific, we will analyze the effect of phase fluctuations on the electron density of states (DOS) of superconducting nanowires.
In order to understand the basic physics behind this effect let us recall that such wires host sound-like plasma modes \cite{Mooij,Buisson} which can be
described in terms of phase fluctuations of the superconducting order parameter. These so-called Mooij-Sch\"on modes can propagate along the wire with the velocity $v=1/\sqrt{\mathcal{L}_{\rm kin}C}$ and interact with electrons inside the wire, thereby forming an effective environment for such electrons and affecting the superconducting DOS.

The structure of the paper is as follows. In Sec. II we define our model and specify the basic formalism employed in our analysis.
In Sec. III we derive the general expression for the quasiclassical electron Green function in superconducting nanowires in the presence of phase fluctuations.
This expression is then employed to evaluate the electron DOS in such nanowires in Sec. IV. Sec. V is devoted to a brief discussion of our key observations.

\section{The model and basic formalism}
Below we will analyze the structure displayed in Fig. 1. A long superconducting wire with sufficiently small cross section $s$ is attached to two big superconducting reservoirs.
As usually, superconducting properties of the system are described by the order parameter field $\Delta (x, t)=|\Delta (x, t)|\exp (i\varphi (x, t))$
which in general depends both on the coordinate along the wire $x$ and on time $t$. The wire parameters are chosen in a way to enable one to disregard all fluctuations of the absolute value of the order parameter which is set to be equal to a constant value $|\Delta (x, t)|=\Delta$ independent of both $x$ and $t$.
As we already discussed above, for this purpose we need to set the dimensionless conductance $g_\xi $ to be large $g_\xi \gg 1$.
On the other hand, we will allow for fluctuations of the phase variable $\varphi (x, t)$ along the wire.

In what follows we will assume our superconducting wire to remain in thermodynamic equilibrium at temperature $T$ well below the superconducting gap,
i.e. $T \ll \Delta$. We will perform our analysis in the most relevant diffusive limit implying that the elastic electron mean free path $\ell$ is
much smaller than the superconducting coherence length $\xi$.

\begin{figure}
\includegraphics[width=0.99\linewidth]{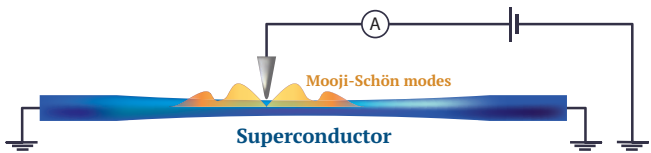}
\caption{(Color online) The system under consideration: A narrow superconducting wire together with a circuit employed for DOS measurements.}
\end{figure}

We will operate with the quasiclassical electron Green function
\begin{equation}
\check{G}(t,t^\prime,x)=\begin{pmatrix}
G^R(t,t^\prime,x) & G^A(t,t^\prime,x)\\
0 & G^K(t,t^\prime,x)
\end{pmatrix},
\label{Gr}
\end{equation}
which has the matrix structure in both Keldysh and Nambu spaces and satisfies the Usadel equation \cite{Usadel,bel}
\begin{equation}
\left[
\partial_t \sigma_3 -i\check{\Delta}+i e\check{V}\sigma_0,\check{G}
\right]
-\frac{D}{2}\hat{\partial}
\left[
\check{G},\hat{\partial}\check{G}
\right]=0
\label{usadel_eq}
\end{equation}
together with the standard normalization condition $\check{G}^2=\check{1}$.

In Eq. (\ref{usadel_eq}) we introduce the covariant spatial derivative
$\hat{\partial}(\dots)=\partial_x(\dots)+ie\left[\check{{A}}_x\sigma_3,(\dots)\right]$, $[a,b]=ab-ba$ denotes the commutator, $V$ and $A$ are the
scalar and vector potentials of the electromagnetic field, $D=v_F\ell /3$ is the diffusion coefficient, $\t_a$ and $\s_a$ (together with the unity matrices $\t_0=\s_0=\hat{1}$) stand for the Pauli matrices respectively in Keldysh and Nambu spaces and $\check{\Delta}$ is the order parameter matrix to be defined below. We also note that all matrix products are understood as convolutions
\begin{equation}
\left(A B\right)(t_1,t_2,x)=\int dt\, A(t_1,t,x)B(t,t_2,x),
\end{equation}
while taking the trace implies integration over both time and space coordinates
\begin{equation}
\tr \,A=\int dt\, dx\,\tr\, A(t,t,x).
\end{equation}

The electron DOS $\nu(E,x)$ is related to the quasiclassical Green function (\ref{Gr}) by means of the equation
\begin{equation}
\nu(E,x)=\nu_0 \tr
\frac{\s_3}{4}\left(
G^R(E,x)-G^A(E,x)
\right),
\end{equation}
where $\nu_0$ stands for DOS in a normal metal at the Fermi level and
\begin{equation}
\G(E,x)=\int d(t-t^\prime) {\rm e}^{iE(t-t^\prime)}\G(t,t^\prime,x).
\end{equation}

It will be convenient for us to perform the rotation in the Keldysh space expressing initial field variables, e.g., the phase of the order
parameter $\fee_{F,B}$ on the forward and backward branches of the Keldysh time contour in terms of their classical and quantum components
$\fee_{\pm}= \left(\fee_F\pm\fee_B\right)/2$. We also define the matrices
\begin{equation}
\check{\fee}=\begin{pmatrix}
\fee_{+} & \fee_-\\
\fee_- & \fee_{+}
\end{pmatrix}
\end{equation}
and
\begin{equation}
\check{\Delta}=\t_0\otimes\begin{pmatrix}
0 & \Delta_{+}\\
-\Delta^*_{+} &0
\end{pmatrix}+\t_1\otimes\begin{pmatrix}
0 & \Delta_{-}\\
-\Delta^*_{-} &0
\end{pmatrix},
\end{equation}
where $\Delta_\pm$ are defined analogously to $\varphi_\pm$.

\section{Green functions in the presence of phase fluctuations}
The task at hand is to average the Green function (\ref{Gr}) over both the fluctuating phase variable $\varphi$ and the electromagnetic field. The latter step is
easily accomplished within the saddle point approximation which allows to directly link the potentials $V$ and $A$  to the phase variable $\fee$ \cite{GZQPS,OGZB}. Employing the gauge transformation
\begin{gather}
e\check{V}\rightarrow \check{\Phi}\equiv e\check{V}+\frac{\dot{\check{\fee}}}{2},
\label{10}\\
e\check{A}_x\rightarrow \check{\A}\equiv e\check{A}_x-\frac{\partial_x \check{\fee}}{2},\\
{\Delta}_{\pm}\rightarrow |\Delta|_{\pm},
\label{12}\end{gather}
we expel the phase of the order parameter from $\Delta (x,t)$ and get
\begin{equation}
\G(t,t^\prime,x)={\rm e}^{\frac{i}{2}\check{\fee}(t,x)\s_3}\Gt(t,t^\prime,x)\nonumber {\rm e}^{-\frac{i}{2}\check{\fee}(t^\prime,x)\s_3},
\label{gauge_transform}
\end{equation}
where $\Gt$ obeys Eq. (\ref{usadel_eq}) combined with Eqs. (\ref{10})-(\ref{12}). It is also necessary to keep in mind that under the condition $g_\xi \gg 1$ adopted here one has $|\Delta|_{+}= \Delta$ and  $|\Delta|_{-}=0$.

As usually, magnetic effects associated with $\A$ remain weak and can be neglected by setting $\A =0$ \cite{GZQPS,OGZB}. Likewise, one can disregard the effects related to weak ($\propto \Phi$) penetration of the fluctuating electric field inside the wire as compared to those caused by the gauge factors in (\ref{gauge_transform}). This conclusion can be drawn from the equation \cite{GZQPS,OGZB}
\begin{equation}
\Phi\sim \dot{\fee} /(4E_C\nu_Fs), \quad E_C=e^2/(2C)
\end{equation}
combined with the observation that the condition $E_C\nu_F s \gg 1$ is usually well satisfied in generic metallic wires.

Thus, we may set $\Gt$ equal to the Green function $\L$ of a uniform superconductor in thermodynamic equilibrium, i.e.
\begin{equation}
\Gt=\L=\begin{pmatrix}
\Lambda^R & \Lambda^K \\
0 & \Lambda^A
\end{pmatrix},
\end{equation}
where
\begin{equation}
\Lambda^R_\epsilon =\frac{1}{\sqrt{(\e+i0)^2-\Delta^2}}
\begin{pmatrix}
\e & \Delta\\
-\Delta & -\e
\end{pmatrix},
\label{lambda_r}
\end{equation}
$\Lambda^A=-\s_3 (\Lambda^R)^\dag \s_3$ and
\begin{equation}
\Lambda^K_\epsilon=\Lambda^R_\epsilon F_\epsilon  - F_\epsilon \Lambda^A_\epsilon,
\quad F_\epsilon =\tanh{\frac{\e}{2T}}.
\label{fermionic_fdt}
\end{equation}
Then we obtain
\begin{gather}
   \G(t,t^\prime,x)\simeq {\rm e}^{ \frac{i}{2}\check{\fee}(t,x)\s_3}\L(t-t^\prime ){\rm e}^{-\frac{i}{2}\check{\fee}(t^\prime,x)\s_3}.
   \label{Gre}
\end{gather}
Here $\L(t-t^\prime)$ is the inverse Fourier transform of $\L_\epsilon$.

What remains is to average the Green function (\ref{Gre}) over all possible phase configurations. This averaging is accomplished
with the aid of the path integral
\begin{equation}
\left\langle \check{G}\right\rangle_\fee (t-t^\prime)=\int D\fee \, \exp\left({iS_{\rm eff}[\fee]}\right)\G(t,t^\prime,x).
\end{equation}
Here $S_{\rm eff}[\fee]$ is the effective action which accounts for phase fluctuations in a superconducting wire. At low energies this action reads \cite{AGZ,GZQPS,OGZB}
\begin{eqnarray}
S_{\rm eff}[\fee]=\frac{C}{4e^2}\tr\left[
\begin{pmatrix}
\fee_{+} & \fee_{-}
\end{pmatrix}\V^{-1}
\begin{pmatrix}
\fee_{+}\\
\fee_{-}
\end{pmatrix}
\right],
\label{Seff}
\end{eqnarray}
where
\begin{equation}
\V=\begin{pmatrix}
\V^K & \V^R\\
\V^A & 0
\end{pmatrix}
\end{equation}
is the equilibrium Keldysh matrix propagator for plasma modes and
\begin{gather}
\V^{R,A}(\omega,k)={\displaystyle\frac{1}{(\omega\pm i0)^2-(kv)^2}},\\
\V^K(\omega,k)=\left(\V^R(\omega,k)-\V^A(\omega,k)\right)\coth{\frac{\omega}{2T}}.
\label{bosonic_fdt}
\end{gather}

\section{Density of states}
Let us now implement the above program and evaluate the electron DOS in superconducting nanowires. Making use of the structure of $\L$ in the Nambu space and performing Gaussian integration, we get
\begin{multline}
\nu(E)=\nu_0\int d(t-t^\prime){\rm e}^{iE(t-t^\prime)}\\\times
\tr\left\langle \frac{\t_3\s_3}{4}{\rm e}^{ \frac{i}{2}\check{\fee}(t,x)\s_3}\L(t-t^\prime){\rm e}^{ -\frac{i}{2}\check{\fee}(t^\prime,x)\s_3}\right\rangle_\fee\\
=\nu_0\int dt\,{\rm e}^{iEt}\tr
\left(\frac{\t_3\s_3}{4}\,\t_a \L(t)\t_b \bB^{ab}(t)\right),\hspace*{0.5cm}
\label{averaging_result}
\end{multline}
where $a,b=\{0,1\}$,
\begin{multline}
\bB(t)=\begin{pmatrix}
\bB^K(t) & \bB^R(t)\\
\bB^A(t) & 0
\end{pmatrix}={\rm e}^{ iE_C(\V^K(t)-\V^K(0))}\\
\times\begin{pmatrix}
\cos\left(E_C(\V^R(t)-\V^A(t))
\right) & i\sin
\left(E_C\V^R(t)
\right)\\
i\sin
\left(E_C\V^A(t)
\right) & 0
\end{pmatrix}\hspace*{0.25cm}
\label{24}
\end{multline}
and
\begin{equation}
\V(t)=\V(t,0)=\int \frac{d\omega dk}{(2\pi)^2}\,{\rm e}^{-i\omega t}\V(\omega,k).
\end{equation}
Note that Eq. (\ref{averaging_result}) accounts for all emission and absorption processes of multiple plasmons in our system via an auxiliary propagator $\bB$.
This propagator obeys the standard causality requirements and satisfies bosonic fluctuation-dissipation theorem (FDT) because plasmons remain in thermodynamic equilibrium, cf. Eq. (\ref{bosonic_fdt}).

Taking the trace in the Keldysh space, employing the FDT relation for the bare Green function and the fluctuation propagator and, finally, evaluating the trace in the Nambu space, from Eq. (\ref{averaging_result}) we obtain
\begin{multline}
\left\langle
\nu
\right\rangle_{\fee}(E)=
\frac{\nu_0}{4}\int dt{\rm e}^{-iEt}\tr
\left(\s_3
\left(
\Lambda^R(t)-\Lambda^A(t)
\right)\bB^K(t)\right.\\
\left. +\s_3\Lambda^K(t)
\left(
\bB^R(t)-\bB^A(t)
\right)
\right)
\\
=\int \frac{d\e}{2\pi}\nu_{BCS}(\e)\bB^K(E-\e)\left(
1+F_\epsilon F_{E-\epsilon}\right),\hspace{0.55cm}
\label{average_dos_formula}
\end{multline}
where $\nu_{BCS}(\e)$ is the BCS density of states in a bulk superconductor.

It is easy to observe that for $\epsilon \gtrsim E+2T$ the combination $1+F_\epsilon F_{E-\epsilon}$ decays as $\propto \exp ((E-\epsilon)/T)$.
Hence, at subgap energies the electron DOS is suppressed by the factor $\sim \exp ((E-\Delta)/T)$ and at $T \to 0$ the superconducting gap $\Delta$ is
not affected by the Mooij-Sch\"on plasmons.

Evaluating $\bB^K$ in Eq. (\ref{24}), one finds
\begin{multline}
\bB^K(t)=\exp \left(-\frac{1}{g} \int\limits_0^{\omega_c} d\omega\,\frac{1-\cos(\omega t)}{\omega}\coth\left(\frac{\omega}{2T}\right)\right)\\
\times\cos\left(\frac{1}{g} \int\limits_0^{\omega_c} d\omega\,\frac{\sin(\omega t)}{\omega}\right).
\label{Bkexpression}
\end{multline}
Here and below we define
$$
\int\limits_{-\omega_c,0}^{\omega_c} d\omega\,(...)=\int\limits_{-\infty,0}^\infty d\omega\,{\rm e}^{\d-\frac
{|\omega|}{\,\,\omega_c}}(...),
$$
where $\omega_c \sim \Delta$ sets the high frequency cutoff which follows naturally from the fact that
the effective action defined in Eqs. (\ref{Seff})-(\ref{bosonic_fdt}) remains applicable only at energies well below the superconducting gap.

It is straightforward to observe that $\bB^K(t=0)=1$ and, hence,
\begin{equation}
\int dE\, \left(\nu(E)-\nu_{BCS}(E)\right)=0.
\end{equation}
This identity implies that phase fluctuations can only redistribute the electron states among different energies
but do not affect the total (energy integrated) DOS.

At low temperatures Eq. (\ref{Bkexpression}) can be evaluated explicitly. We obtain
\begin{multline}
\bB^K(t)=\left(\frac{\sinh(\pi T t)}{\pi T t}\sqrt{1+(\omega_c t)^2}\right)^{-1/g}\\
\times\cos\left(\frac{\arctan(\omega_c t)}{g}\right).
\end{multline}
In order to recover $\bB^K(\omega)$ it is convenient to express it in terms of the Matsubara propagator for the phase fluctuations
continued analytically to the complex plane. For this purpose let us define
\begin{equation}
\bB^K(t)=\frac{1}{2}\sum_\pm
{\rm e}^{{-\left(\D(0)-\D(t\pm i0)\right)/g}},
\end{equation}
where we introduced the propagator
\begin{equation}
\D(t\pm i0)=\int_{-\omega_c}^{\omega_c} \frac{d\omega}{2\omega}{\rm e}^{-i\omega t}\left(
\coth\left(\frac{\omega}{2T}\right)\mp 1
\right).
\end{equation}
This propagator is periodic in imaginary time and has cuts at $\mathrm{Im}\ t =\beta n$ with $\beta=1/T$ and $n\in \mathbb{Z}$. Shifting the integration contour, one obtains
\begin{equation}
\bB^K(\omega)=\cosh\left(
\frac{\beta \omega}{2}
\right)\int dt\, \mathrm{e}^{-i\omega t}\,\bB^K\left(t+\frac{i\beta}{2}\right),
\end{equation}
where
\begin{equation}
\bB^K\left(t+\frac{i\beta}{2}\right)=\exp \left(-\frac{1}{g} \int\limits_0^{\omega_c} \frac{d\omega}{\omega}\frac{\cosh\left(\frac{\omega}{2T}\right)-\cos(\omega t)}{\sinh\left(\frac{\omega}{2T}\right)}\right)
\end{equation}
These integrals can easily be evaluated with the result
\begin{multline}
\bB^K(\omega)\simeq \cosh\left(
\frac{\beta \omega}{2}
\right)\left(\frac{2\pi T}{\omega_c}\right)^{1/g}
\frac{\left|\Gamma\left(\d\frac{1}{2g}+\frac{i\omega}{2\pi T}\right)\right|^2}{2\pi T \Gamma(1/g)},
\label{34}
\end{multline}
where $\omega$ is supposed to be well below the superconducting gap $\Delta$. For $\omega\ll T$ Eq. (\ref{34}) reduces to
\begin{equation}
\bB^K(\omega)\simeq \frac{1}{g\omega_c}\left(\frac{2\pi T}{\omega_c}\right)^{1/g}\frac{2\pi T}{\omega^2+(\pi T/g)^2},
\end{equation}
whereas at higher frequencies $T \ll \omega\ll \Delta$ we find
\begin{equation}
\bB^K(\omega)\simeq \frac{\pi}{\omega_c\Gamma(1/g)}\left(\frac{\omega}{\omega_c}\right)^{1/g-1}.
\end{equation}

Making use of the above expressions, at energies in the vicinity of the superconducting gap $\Delta$ we recover the following result for the electron DOS:
\begin{multline}
\nu(\Delta+\omega)=\frac{\nu_0\sqrt{\Delta}}{\sqrt{2}}\left(\frac{2\pi T}{\Delta}\right)^{1/g}\sum\limits_{k=0}^\infty \frac{\Gamma(k+1/g)}{k!\Gamma(1/g)}\\
\times{\rm Re}\left(
\frac{{\rm e}^{-\frac{i\pi}{2g}}}{\sqrt{\omega+2i\pi T(\frac{1}{2g}+k)}}
\right).
\end{multline}

\begin{figure}
\includegraphics[width=0.99\linewidth]{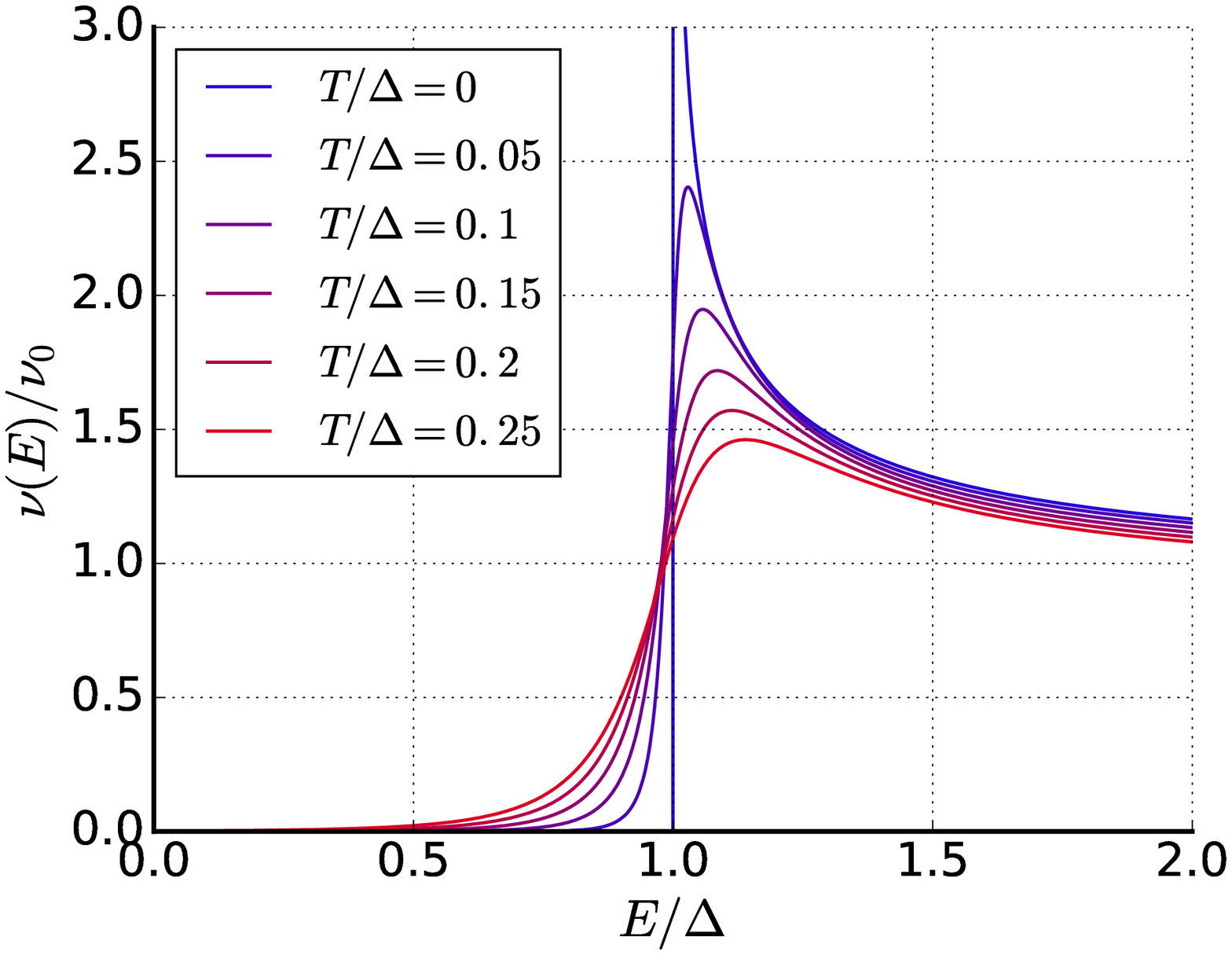}
\includegraphics[width=0.99\linewidth]{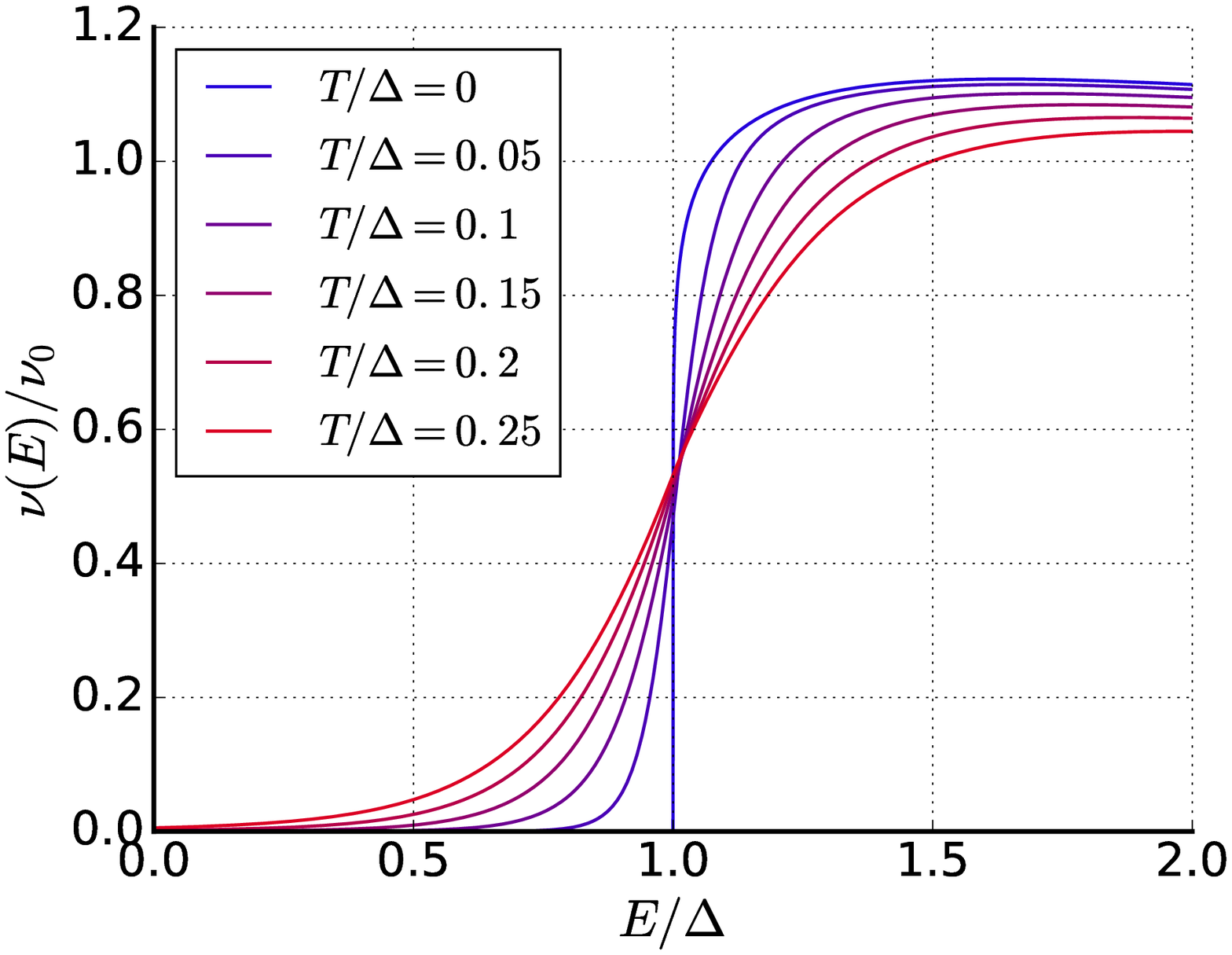}
\caption{(Color online) The normalized energy dependent electron density of states $\nu(E)/\nu_0$ for superconducting nanowires at different temperatures and two
 values of $g=5$ (top) and $g=1.67$ (bottom). The energy $E$ and temperature $T$ are expressed in units of $\Delta$.}
\end{figure}

The energy dependent density of states $\nu (E)$ for superconducting nanowires in the presence of phase fluctuations is also displayed in Fig. 2
at different temperatures and two different values of the dimensionless conductance $g$. One observes that at any nonzero $T$ the BCS singularity
at $E \to \Delta$ is smeared due to interactions between electrons and Mooij-Sch\"on plasmons. For the same reason, as we already indicated above, the electron DOS at subgap
energies $0<E<\Delta$ remains non-zero at any non-zero $T$, i.e.
\begin{equation}
\nu (E) \propto \exp ((E-\Delta)/T).
\label{38}
\end{equation}

We also point out a qualitative difference in the energy dependence of DOS displayed in top and bottom panels of Fig. 2 at energies slightly above the gap. While at bigger values of $g$  the function $\nu (E)$ demonstrates a non-monotonous behavior at such energies (top panel), at smaller $g$ DOS decreases monotonously with decreasing energy at all $E$ not far from the gap (bottom panel). In the zero temperature limit $T \to 0$ and for $E-\Delta \ll \Delta$ we obtain
\begin{equation}
\nu(E)\simeq \frac{\nu_0 \sqrt{\pi} \theta(E-\Delta)}{\sqrt{2}\Gamma(\frac{1}{2}+\frac{1}{g})}\left(
\frac{E-\Delta}{\Delta}
\right)^{\frac{1}{g}-
\frac{1}{2}}.
\label{DOST0}
\end{equation}
We observe that while at $E<\Delta$ the electron DOS (\ref{DOST0}) vanishes at all values of $g$, the behavior of $\nu (E)$ at overgap energies is markedly different depending on the dimensionless conductance $g$. For $g > 2$ (i.e. for relatively thicker wires) the DOS singularity at $E \to \Delta$ survives
though becoming progressively weaker with decreasing $g$. On the other hand,  at $g\leq 2$ (corresponding to relatively thinner wires)
the DOS singularity vanishes completely due to intensive phase fluctuations and $\nu (E)$ tends to zero at $E \to \Delta$ as a power law (\ref{DOST0}).
This behavior is also illustrated in Fig. 3.

\begin{figure}
\includegraphics[width=0.99\linewidth]{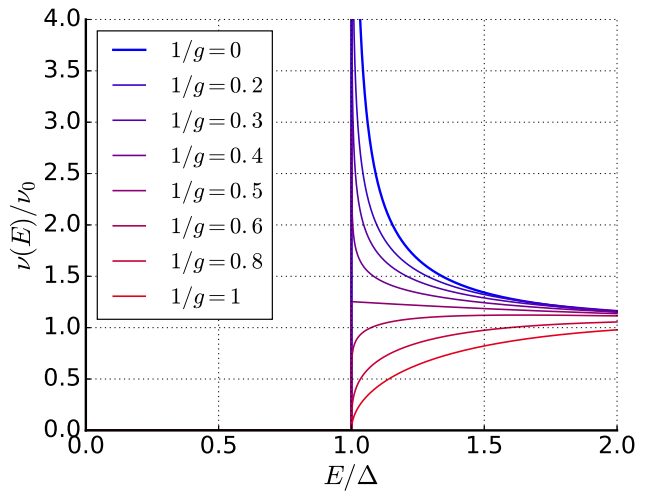}
\caption{(Color online) The same as in Fig. 2 at $T=0$ and different values of $g$.}
\end{figure}

\section{Discussion}
In this paper we argued that fluctuations of the phase of the order parameter may significantly affect low temperature properties of superconducting
nanowires. While the dramatic effect of spin-wave-like fluctuations on long-range phase coherence in quasi-one-dimensional systems is well known
for a long time \cite{HMW}, here we demonstrated that {\it local} properties of superconducting nanowires, such as the electron density of states,
can also be sensitive to phase fluctuations. We deliberately chose the wire parameters in a way to minimize fluctuations of
the absolute value of the order parameter
and specifically addressed the effect of small phase fluctuations associated with low energy sound-like plasma modes propagating along the wire.
These Mooij-Sch\"on plasmons form an effective quantum dissipative environment for electrons inside the wire. Previously various ground state properties
of superconducing nanorings affected by such an environment were explored by several authors \cite{HG,SZ13}. Here we adopted
a physically similar standpoint in order to investigate the behavior of the electron DOS in long superconducting nanowires.

The coupling strength between electrons and the effective plasmon environment is controlled by the dimensionless parameter $g$ representing the ratio between the quantum resistance unit $R_q$ and the wire impedance $Z_{\rm w}$. Provided $g \gg 1$, i.e. the impedance $Z_{\rm w}$ remains much smaller
than $R_q$, phase fluctuations weakly affect the electron DOS except in the immediate vicinity of the superconducting gap $\Delta$. For larger values
$Z_{\rm w} \sim R_q$ the effect of phase fluctuations becomes strong and should be treated non-perturbatively in $1/g$ at all energies. Another important
parameter is temperature which is restricted here to be sufficiently low $T \ll \Delta$.

Our analysis demonstrates that at any nonzero $T$ the electron DOS depends on temperature and substantially deviates from that derived from the standard BCS theory. In particular, at $T >0$ the BCS square-root singularity in DOS at $E=\Delta$ gets totally smeared and $\nu (E)$ differs from zero also at subgap energies, cf. Eq. (\ref{38}).  This behavior can be interpreted in terms of a depairing effect due to the interaction between electrons and Mooij-Sch\"on plasmons. We also note that our results are consistent with the phenomenological Dynes formula \cite{dynes78}
\begin{equation}
\nu (E)\simeq \nu_0{\rm Re}\left(\frac{E+i\Gamma }{\sqrt{(E+i\Gamma )^2-\Delta^2}}\right)
\label{Dynes}
\end{equation}
describing smearing of the BCS singularity in DOS in the immediate vicinity of the superconducting gap.

At $T=0$ and subgap energies the electron DOS vanishes as in the BCS theory, while the BCS singularity in DOS at $E \to \Delta$ becomes weaker for any finite $g >2$ and eventually disappears for $g \leq 2$. Thus, we conclude that even in the absence of fluctuations of the absolute value of the order parameter $|\Delta |$ quantum fluctuations of its phase $\varphi$ may result in qualitative modifications of the ground state properties of quasi-one-dimensional superconducting wires.

The local electron DOS in superconducting nanowires can be probed in a standard manner by performing a tunneling experiment, as it is also illustrated in
Fig. 1. Attaching a normal or superconducting electrode to our wire and measuring the differential conductance of the corresponding tunnel junction
one gets a direct access to the energy dependent electron DOS of a superconducting nanowire. E.g., in the case of a normal electrode at $T \to 0$ and $eV > \Delta$ one finds
\begin{equation}
dI/dV \propto \nu (eV) \propto (V -\Delta /e)^{\frac{1}{g}-
\frac{1}{2}}.
\label{CA}
\end{equation}
This power law dependence of the differential conductance resembles one encountered in small normal tunnel junctions at low voltages $dI/dV \propto V^{2/g_N}$ \cite{PZ88}, where $g_N$ is the dimensionless conductance of normal leads. In fact, both the dependence (\ref{CA}) and the zero bias anomaly
in normal metallic junctions \cite{PZ88} are caused by Coulomb interaction and are controlled by the impedance of the corresponding effective electromagnetic environment.

Finally, we remark that in superconducting nanowires with not too large values of $g_\xi$ it is also necessary to account for quantum fluctuations of the absolute value of the order parameter $|\Delta |$. Such fluctuations combined with those of the phase $\varphi$ result in a reduction of the superconducting gap \cite{GZTAPS} and cause Berezinskii-Kosterlitz-Thouless-like (superconductor-insulator) quantum phase transition for QPS \cite{ZGOZ} at $\lambda \equiv g/8 =2$. A complete analysis of quantum fluctuations and their impact on the electron DOS in superconducting nanowires should include all the effects controlled by both parameters $g$ and $g_\xi$. This analysis will be worked out elsewhere.

\vspace{0.5cm}

\centerline{\bf Acknowledgements}
We would like to thank K.Yu. Arutyunov for encouragement and useful discussions. This work was supported by the Russian Science Foundation under grant No. 16-12-10521.


\begin{references}
\bibitem{AGZ} K.Yu. Arutyunov, D.S. Golubev, and A.D. Zaikin, Phys. Rep. {\bf 464}, 1 (2008).
\bibitem{LV} A.I. Larkin and A.A. Varlamov, {\it Theory of fluctuations in superconductors} (Clarendon, Oxford, 2005).
\bibitem{ZGOZ} A.D. Zaikin, D.S. Golubev, A. van Otterlo, and G.T. Zimanyi, Phys. Rev. Lett. {\bf 78}, 1552 (1997).
\bibitem{GZQPS} D.S. Golubev and A.D. Zaikin, Phys. Rev. B {\bf 64}, 014504 (2001).
\bibitem{BT} A. Bezryadin, C.N. Lau, and M. Tinkham, Nature {\bf 404}, 971 (2000).
\bibitem{Lau} C.N. Lau, N. Markovic, M. Bockrath, A. Bezryadin, and M. Tinkham, Phys. Rev. Lett. {\bf 87},  217003 (2001).
\bibitem{Zgi08} M. Zgirski, K.P. Riikonen, V. Touboltsev, and K.Y. Arutyunov, Phys. Rev. B {\bf 77},  054508 (2008).
\bibitem{SZ16} A.G. Semenov and A.D. Zaikin, Phys. Rev. B {\bf 94}, 014512 (2016).
\bibitem{GZTAPS} D.S. Golubev and A.D. Zaikin, Phys. Rev. B {\bf 78}, 144502 (2008).
\bibitem{Mooij} J.E. Mooij and G. Sch\"on, Phys. Rev. Lett. {\bf 55}, 114 (1985).
\bibitem{Buisson} B. Camarota, F. Parage, F. Balestro, P. Delsing, and O. Buisson, Phys. Rev. Lett. {\bf 86}, 480 (2001).
\bibitem{Usadel} K.D. Usadel, Phys. Rev. Lett. \textbf{25}, 507, (1970).
\bibitem{bel} W. Belzig, F. Wilhelm, C. Bruder, G. Sch\"on, and A.D. Zaikin, Superlatt. Microstruct. \textbf{25}, 1251 (1999).
\bibitem{OGZB} A. van Otterlo, D.S.Golubev, A.D.Zaikin, and G.Blatter, Eur. Phys. J. B {\bf 10}, 131 (1999).
\bibitem{HMW} P.C. Hohenberg, Phys. Rev. {\bf 158}, 383 (1967);
  N.D. Mermin and H. Wagner, Phys. Rev. Lett. {\bf 17}, 1133 (1966).
\bibitem{HG} F.W.J. Hekking and L.I. Glazman,  Phys. Rev. B {\bf 55}, 6551 (1997).
\bibitem{SZ13} A.G. Semenov and A.D. Zaikin, Phys. Rev. B {\bf 88}, 054505 (2013).
\bibitem{dynes78} R.C. Dynes, V. Narayanamurti and J.P. Garno, Phys. Rev. Lett. {\bf 41}, 1509 (1978).
\bibitem{PZ88} S.V. Panyukov and A.D. Zaikin, J. Low Temp. Phys. {\bf 73}, 1 (1988).




\end{references}
\end{document}